%% file: submit.tex
\documentclass[aps,prd,twocolumn,superscriptaddress,preprintnumbers,floatfix,nofootinbib,notitlepage,showkeys,showpacs]{revtex4-1}

\usepackage[utf8x]{inputenc}
\usepackage{graphicx}
\usepackage{hyperref}
\usepackage{latexsym}
\usepackage{amsmath}
\usepackage{amssymb}
\usepackage{bbm}

\usepackage{ulem}
\usepackage{pdfsync}
\usepackage{epsfig}
\usepackage{epstopdf}

\newcommand{\Tr}{\textrm{Tr}\,}
\newcommand{\tr}[1]{\,\text{Tr}\!\left ( #1 \right )}

\newcommand{\ev}[1]{\left \langle #1 \right \rangle}
\newcommand{\bs}[1]{{\bf #1}} 

\newcommand{\csw}{c_{\rm{sw}}}
\newcommand{\eq}[1]{Eq.~(\ref{#1})}

\newcommand{\Dslash}{\ensuremath \raisebox{0.025cm}{\slash}\hspace{-0.27cm} D}
\newcommand{\be}{\begin{equation}}
\newcommand{\ee}{\end{equation}}
\newcommand{\bea}{\begin{eqnarray}} 
\newcommand{\eea}{\end{eqnarray}}

\newcommand{\bmp}{\noindent\begin{minipage}{16cm}}
\newcommand{\emp}{\end{minipage}\vskip 7mm} 
\def\lsim{\mathrel{\raise.3ex\hbox{$<$\kern-.75em\lower1ex\hbox{$\sim$}}}}
\def\gsim{\mathrel{\raise.3ex\hbox{$>$\kern-.75em\lower1ex\hbox{$\sim$}}}}

\usepackage{color}

\newcommand{\kr}[1]{#1}
\newcommand{\krsout}[1]{}

\newcommand{\jr}[1]{#1}
\newcommand{\jrsout}[1]{}

\newcommand{\Proj}{{\rm{P}}}

\newcommand{\intron}[1]{}

\begin{document}

\title{Running coupling in SU(2) gauge theory with two adjoint fermions}
  

\author{Jarno Rantaharju}
\email{rantaharju@cp3.sdu.dk}
\affiliation{CP-Origins \& IMADA, Campusvej 55, DK-5230 Odense M, Denmark\\ and RIKEN Advanced Institute for Computational Science, Kobe, Hyogo 650-0047, Japan}
\author{Teemu Rantalaiho}
\email{teemu.rantalaiho@helsinki.fi}
\author{Kari Rummukainen}
\email{kari.rummukainen@helsinki.fi}
\author{Kimmo Tuominen}
\email{kimmo.i.tuominen@helsinki.fi}
\affiliation{Department of Physics and Helsinki Institute of Physics,
 P.O.Box 64, FI-00014 University of Helsinki, Finland}

%

\begin{abstract}
We study SU(2) gauge theory with two Dirac fermions in the adjoint representation of the gauge group on the lattice. Using clover improved Wilson fermion action with hypercubic truncated stout smearing we perform simulations at larger coupling than earlier. We measure the evolution of the coupling constant using the step scaling method with Schr\"odinger functional \jr{and study the remaining discretization effects}. \kr{At weak coupling we observe significant discretization effects, which make it difficult to obtain a fully controlled continuum limit. Nevertheless, the data remains consistent with the} existence of a fixed point in the interval $2.2\lsim g^{\ast 2}\lsim 3$. We also measure the anomalous dimension and find its value at the fixed point is $\gamma^\ast\simeq 0.2\pm 0.03$.
\end{abstract}

\preprint{CP3-Origins-2015-040 DNRF90, DIAS-2015-40, HIP-2015-33/TH}



\maketitle


\section{Introduction}

Quantitative determination of the vacuum phase of an SU($N_c$) gauge theory with massless fermions 
as a function of the number of colors, $N_c$, flavors, $N_f$ and fermion representations
provides a challenge for solving nonperturbative strong dynamics. 
Of particular interest is the location of the  conformal window, i.e. the range of values of $N_f$ for given $N_c$ and fermion representation, where the theory has a nontrivial infrared fixed point (IRFP) governing the large distance behavior of the theory.

To concretise, consider the two-loop beta function,
\be
\beta(g^2)=\frac{dg^2}{d\log\mu^2}=-\frac{\beta_1}{16\pi^2}g^4-\frac{\beta_2}{(16\pi^2)^2} g^6,
\ee
for a fixed value of $N_c$ and massless quarks transforming under some representation ${\cal R}$ of SU($N_c$).
First, at small enough $N_f$ the physics is QCD-like and $\beta(g^2)$ is negative for all values of the coupling and at low energy strong SU($N_c$) dynamics induces formation of quark-antiquark condensate breaking the chiral symmetry. On the other hand, the asymptotic freedom is lost above $N_f=N_{f,0}$, as determined by the vanishing of the one-loop coefficient of the beta function, $\beta_1(N_c,N_{f,0})=0$. 
In the region directly below this upper boundary, the theory is weakly coupled and one can establish the existence of a nontrivial IRFP rigorously by perturbation theory \cite{Banks:1981nn}. 
However, when $N_f$ is decreased significantly from $N_{f,0}$, the fixed point shifts towards larger couplings, and the spontaneous formation of chiral condensate may occur inhibiting the flow into the IRFP implied by the two-loop beta function. The value $N_{f,{\rm{crit}}}$ where the transition from IRFP behavior to spontaneous chiral symmetry breaking takes place defines the location of the lower boundary of the conformal window, and must be determined by nonperturbative methods.

While the studies of the phase diagrams of gauge theories in general are motivated by intrinsic interest 
into strong dynamics, they also have applications in constructing models beyond the Standard 
Model. A prime example are the technicolor theories, where the electroweak symmetry is broken by a 
spontaneous chiral symmetry breaking of a strong interaction 
\cite{TC,Eichten:1979ah,Hill:2002ap,Sannino:2008ha}. 
Over the last few years there has been 
significant interest in the exploration of quantum gauge theories with matter in fundamental or higher representation. Using various approximations, the location of the conformal window has been estimated and possible candidates for beyond Standard Model theories have been identified \cite{Sannino:2004qp}. 
Lattice simulations provide the only first principle method for a precision analysis of the non-perturbative properties of these theories.

In this work we study the SU(2) gauge field theory coupled to two massless fermions in the adjoint representation.\footnote{%
In a related work, the existence of the infrared fixed point in SU(2) gauge theory with different numbers of fermions in the fundamental representation of the gauge group has 
 been recently studied in 
\cite{Bursa:2010xr,Ohki:2010sr,Voronov:2013ba,Karavirta:2011zg,Voronov:2012qx,Hayakawa:2013maa,Hayakawa:2013yfa,Appelquist:2013pqa,Matsufuru:2014uea,Hietanen:2014xca,Athenodorou:2014eua}.}
The lattice studies of this model were initiated in \cite{Catterall:2007yx}, and
the first large scale simulations providing evidence for the existence of an IRFP were reported in \cite{Hietanen:2008mr,Hietanen:2009az}. These results have since then been confirmed by 
several studies of different collaborations \cite{DelDebbio:2008zf,Catterall:2008qk,
    Bursa:2009we,DelDebbio:2009fd,DelDebbio:2010hx,
    DelDebbio:2010hu,Bursa:2011ru,DeGrand:2011qd, Patella:2012da,
    Giedt:2012rj,Rantaharju:2013gz,Rantaharju:2013bva,DelDebbio:2015byq}.
Even though all studies so far favor the existence of an IRFP in this theory, the results should be
interpreted carefully as the slow renormalization group evolution is masked by the discretization effects.
\kr{Especially the evolution of the coupling constant as a function of the energy scale is not yet
known at a fully satisfactory level.}

Implementation of the improved Wilson fermion into these studies was undertaken in 
\cite{Karavirta:2011mv}. Here we furthermore use hypercubic stout (HEX) 
smearing \cite{Capitani:2006ni} \kr{in order to further reduce} the discretization effects. \kr{Similar} methods have been 
successfully applied to reduction of lattice artifacts in QCD simulations.
\kr{We also extend} the smearing to the gauge action.  This allows us to run simulations at stronger couplings, \kr{which is necessary in order to reach the fixed point.}
We measure the running coupling using the Schr\"odinger functional method,
\kr{and while we do not have full control of the continuum limit, the existence of a non-trivial infrared
fixed point is clear.  The results from the largest volumes (smallest lattice spacings) indicate that 
the IRFP is}
close to $g^2\simeq 2$.  
The result is in \kr{overall} agreement with previous studies.

In addition to the existence of the IRFP, the obvious quantities of interest are the scheme independent values of physical observables at the fixed point. These include the slope of the beta-function 
and the anomalous dimension $\gamma$ of the quark mass operator $\bar{\psi}\psi$ which
determines the running of the quark mass as
\be
\mu\frac{dm(\mu)}{d\mu}=-\gamma(g^2)m(\mu).
\ee
The anomalous dimension $\gamma$ is phenomenologically interesting for extended technicolor model building, where the fermion masses are produced by the technicolor symmetry breaking.
The mass anomalous dimension $\gamma^\ast$ of a quasi stable IRFP together with the 
running of the coupling determines the physical fermion masses.
We measure the mass anomalous dimension in our simulations and find a relatively small value 
$\gamma^\ast\simeq 0.2$ at the fixed point.   

The paper is organized as follows: In section \ref{model} we introduce the details of the lattice model we use. In sections \ref{coupling} and \ref{gamma} we discuss the running coupling and the anomalous dimension respectively and present the results obtained from the simulations. In section \ref{checkout} we present our conclusions and outlook.

\section{The lattice model}
\label{model}

The model is defined by the action
\begin{equation}
 S = S_G + S_F,
\end{equation}
where $S_G$ is a partially smeared Wilson plaquette action 
and $S_F$ is the clover improved Wilson fermion action with smeared gauge links.
We use hypercubic truncated stout smearing (HEX smearing) \cite{Capitani:2006ni},
which helps to reduce the discretization errors and allows simulations at larger couplings
than unsmeared action does.

The smeared links are calculated in three sequential stout smearing steps,
each limited to the directions that are orthogonal to those in the previous steps:
{\small 
\begin{align*}
& \overline V_{x,\mu;\nu,\rho} = \Proj \big( \frac{\alpha_3}{2}\sum_{\pm\eta\neq\mu,\nu,\rho} U_{x,\eta} U_{x+\hat\eta,\mu} U^\dagger_{x+\hat\mu,\eta} U^\dagger_{x,\mu} \big) U_{x,\mu}, 
\\
& \tilde V_{x,\mu;\nu} = \Proj \big( \frac{\alpha_2}{4}\sum_{\pm\rho\neq\mu,\nu} \overline V_{x,\rho;\nu,\mu} \overline  V_{x+\hat\rho,\mu;\nu,\rho} \overline V^\dagger_{x+\hat\mu,\rho;\mu,\nu} U^\dagger_{x,\mu} \big) U_{x,\mu}, 
\\
& V_{x,\mu} = \Proj \big( \frac{\alpha_1}{6}\sum_{\pm\nu\neq\mu} \tilde V_{x,\nu;\mu} \tilde V_{x+\hat\nu,\mu;\nu} \tilde V^\dagger_{x+\hat\mu,\nu;\mu} U^\dagger_{x,\mu} \big) U_{x,\mu}
\end{align*}
}
where 
\begin{equation}
 \Proj (U) = \exp\left[ U - U^\dagger - \frac12 \Tr(U -U^\dagger)\right]
\end{equation}
projects the argument to SU(2), and the convention $U_{x,-\mu} = U^\dagger_{x-\hat\mu,\mu}$ is used.

The smearing parameters $\alpha_1$, $\alpha_2$ and $\alpha_3$ were determined by 
maximizing the expectation value of the smeared plaquette
$
 P=\ev{\tr{V_{\mu\nu}(x)}}
$
in simulations with 500 trajectories at $L/a=6$, $\beta=3$ and $\kappa = 0.126$.
This yields the values 
$\alpha_1 = 0.78$, $\alpha_2 = 0.61$ and $\alpha_3 = 0.35$, which are 
close to the standard tree-level values \cite{Capitani:2006ni}.

The gauge action is a mixture of single-plaquette Wilson actions with smeared and unsmeared gauge links:
\begin{align}
  S_G
  &= \beta_L \sum_{x;\mu<\nu} (1-c_g) {\mathcal L}_{x,\mu\nu}(U) + c_g {\mathcal L}_{x,\mu\nu}(V) 
   \label{plaqaction}\\
  {\mathcal L}_{x,\mu\nu}(U) &= 1 - \frac12 \Tr 
 \left [U_{x,\mu} U_{x+\hat\mu,\nu} U^\dagger_{x+\hat\nu,\mu} U^\dagger_{x,\nu} \right] ,
\nonumber
\end{align}
where $\beta_L = 4/g_0^2$.
Using partially smeared action enables us to run simulations at stronger physical couplings, as was observed in \cite{DeGrand:2011qd}.  The properties of the gauge action are not sensitive to the precise value of $c_g$, and for concreteness we choose here $c_g=0.5$. 

The fermions belong to the adjoint (3-dimensional) representation of SU(2).  
We use the Wilson-clover fermion action
\begin{align*}
  S_F
  & = a^4\sum_x \bigg [
  \bar{\psi}(x) ( i\, \Dslash_W + m_0 )
  \psi(x)  \nonumber \\
 & + a \csw \bar\psi(x)\frac{i}{4}\sigma_{\mu\nu}
  F_{\mu\nu}(x)\psi(x) \bigg ],
\end{align*}
where $\Dslash_W$ is the standard Wilson Dirac operator.  The gauge link
matrices appearing in $S_f$ are in the adjoint
representation, which are constructed from the smeared matrices $V_{x,\mu}$ as follows:
\begin{equation}
  \tilde U^{a,b}_{x, \mu} =
  \frac12 \Tr[ \sigma^a V_{x,\mu} \sigma^b V^\dagger_{x,\mu}].
\end{equation}
The full action is conventionally parametrized in terms of the bare coupling $\beta_L=4/g_0^2$, the hopping parameter $\kappa = 1/(2m_0 + 8)$ and the 
Sheikholeslami-Wohlert clover coefficient $\csw$.  We use the tree-level clover coefficient $\csw=1$, which is expected to be a good approximation 
with smeared gauge links \cite{Capitani:2006ni,Shamir:2010cq,DeGrand:2011qd}.
We verified the validity of this assumption by measuring the non-perturbative clover coefficient at small volume using the Schr\"odinger functional tuning method \cite{Luscher:1993gh}.  We find results consistent with the tree-level value even at small values of $\beta_L$.

With unsmeared Wilson fermions this model exhibits a lattice bulk phase transition at large bare coupling, see e.g. \cite{Hietanen:2008mr}.
Such \kr{a} transition is generally signaled by a discontinuity in both the plaquette expectation value and the quark mass with respect to $\kappa$. Along the critical line $\beta_L(\kappa_c)$, where the PCAC quark mass vanishes, towards larger bare couplings, this discontinuity borders the strong coupling region where zero quark mass cannot be reached.  Consequently, in this strong coupling region physical results are not expected.
The utility of the smearing of the fermion and gauge actions is that it moves this bulk transition to larger couplings, expanding the range of parameter values available for measurements.

We measure both the anomalous dimension and the running coupling with the 
Schr\"odinger functional method
\cite{Luscher:1992an,Luscher:1992ny,Luscher:1993gh,DellaMorte:2004bc}. 
We consider a lattice of linear dimension $L$, whose volume $V = L^4 = (Na)^4$.
The spatial boundary conditions for the gauge and fermion fields are periodic, while
the spatial components of the gauge link matrices at timeslices $t=0$ and $t=L$ 
are set to constant values, described in detail in the next section.  
The fermion fields vanish at $t=0,L$ timeslices.  
These boundary conditions remove the fermion zero modes and allow simulations at 
vanishing physical quark masses, which we use here in all of our production runs.

The Wilson fermion action breaks chiral symmetry and requires additive renormalization of the quark mass. Thus, in order to simulate massless theory, we need to determine the critical bare mass (or $\kappa_c(\beta_L)$) where the physical quark mass vanishes.  
We define the quark mass $M$ 
through the lattice PCAC relation \cite{Luscher:1996vw}
\begin{align}
aM(t) &= \frac14 \frac{(\partial_t^* + \partial_t) f_A(t)}{f_P(t)}\\
    &= \frac14\frac{f_A(t+a)-f_A(t-a)}{f_P(t)}
\end{align}
and we define $\kappa_c$ as the value of the parameter $\kappa$ where $M(t=L/2)$ vanishes.
The pseudoscalar current and density correlation functions are
\begin{align}
  f_A(t) &= \frac{-a^6}{3L^6}\sum_{\bs y,\bs z} 
  A_0^a(\bs x,t)
  \,\bar\zeta(\bs y)\gamma_5\frac12\sigma^a\zeta(\bs z)\rangle 
  \label{fa}
  \\
  f_P(t) &= \frac{-a^6}{3L^6}\sum_{\bs y,\bs z} 
  P^a(\bs x, t)
  \,\bar\zeta(\bs y)\gamma_5\frac12\sigma^a\zeta(\bs z)\rangle,
  \label{fp}
\end{align}
where $\zeta$ and $\bar\zeta$ are boundary quark sources at $t=0$, and the axial current and
density can be expressed as 
\begin{align}
  A_\mu^a(x) &= \bar\psi(x)\gamma_\mu\gamma_5 \frac12\sigma^a \psi(x) \\
  P^a(x)     &= \bar\psi(x)\gamma_5 \frac12\sigma^a \psi(x) 
\end{align}
Here $\sigma^a$ is a Pauli matrix acting on the flavour indices of the quark fields.

To find $\kappa_c$ we measure the mass at 3 to 7 values of $\kappa$ on lattices of size $L/a=16$ and interpolate to find where the mass is zero.
The values of $\kappa_c$ used in the simulations are given in table \ref{table:kappac}.
We have also investigated the mass dependence of the measured coupling by reweighting it to the value of $\kappa$ where the mass is zero on the largest lattice, $L=20$. However, this reweighting has negligible effect on all our measurements and we only show the unreweighted data.  We observe no sign of a bulk first order transition even at strongest lattice couplings.

\begin{table}
\centering
\begin{tabular}{llll}
\hline
 $\beta_L$ & $\kappa_c$ & $aM(L/2)$ & $N_{traj}$    \\ \hline
\hline
 8    & 0.125842 & -4(1)e-6  & 118557 \\   
 6    & 0.126251 & -8.0(4)e-5  & 72865 \\ 
 5    & 0.126647 & -1.04(3)e-4  & 134378 \\ 
 4    & 0.127352 & 9.9(4)e-5  & 145775 \\ 
 2    & 0.132309 & 1.3(1)e-4   & 151203 \\ 
 1.5  & 0.136362 & -3.3(2)e-4   & 191039 \\ 
 1.3  & 0.13903  & -9.8(3)e-4  & 170864 \\ 
 1.2  & 0.14073  & -9.5(2)e-4  & 158828 \\ 
 1.1  & 0.142812 & -1.83(4)e-3  & 170602 \\ 
 1.05 & 0.14395  & -4.22(4)e-3  & 128207 \\
 1    & 0.145344 & -6.4(1)e-3 & 35837 \\ 
\hline
\end{tabular}
\caption {
Parameter $\kappa$ used in the simulations and the PCAC mass at each $\beta_L=4/g_0^2$ and the number of measurements performed on the largest lattice.}
\label{table:kappac}
\end{table}

We note that in addition to the clover term, there are order $a$ improvement terms that can be added
to the action at the timelike boundaries of the lattice \cite{Karavirta:2010ef,Karavirta:2011mv}
and to the axial current correlator $f_A$ \cite{Luscher:1996vw}.
Since we have chosen to use the tree-level value for the clover coefficient $\csw$, improving the step scaling function only to the first order in $g^2$, we have consistently chosen to leave these improvements to the tree-level, where they have no effect. 

The simulations are done using the hybrid Monte Carlo (HMC) algorithm with the 2nd order Omelyan integrator \cite{Omelyan,Takaishi:2005tz}
and the chronological initial condition for the fermion matrix inversion \cite{Brower:1995vx}.
The length of the trajectory is fixed to 2 units and the step size is tuned so that the acceptance rate is at least 80\%.
The measurements are taken after every trajectory and the number of trajectories in each simulation varies up to 200,000.

The fermion matrix inversion is acclelerated using the Hasenbusch method 
on lattices of sizes of $L/a=12$ and larger \cite{Hasenbusch:2001xh,Hasenbusch:2002ai}.
The intermediate Hasenbusch mass parameter is chosen to be $m_0= \sqrt[4]{ \lambda_L \lambda_l } $,
where $\lambda_L$ and $\lambda_l$ are estimates of the largest and the smallest eigenvalue
of the two-flavor fermion matrix $M^\dagger M$ \cite{Hasenbusch:2002ai}.
The eigenvalues are measured from short runs with each $\beta$ and $L$.
For the largest lattices, $L/a=20$ and $24$, we split the fermion matrix into three parts and choose the shifts as $m_0 = \sqrt[6]{ \lambda_L^2 \lambda_l } $ and $m_1= \sqrt[6]{ \lambda_L \lambda_l^2 } $.

\section{Evolution of the coupling constant}
\label{coupling}

The Schr\"odinger functional method for measuring the coupling constant is 
based on a background field induced by boundary conditions.
Explicitly, the spatial gauge link matrix boundary conditions are 
\begin{align}
U_i(\bs x,t=0) &= e^{-i\eta \sigma_3 a/L},
\label{sfboundary1}\\
U_i(\bs x,t=L) &= e^{-i(\pi - \eta) \sigma_3 a/L}
\label{sfboundary2}
\end{align}
with $\eta=0.25\pi$.
The fermion fields are set to zero at the temporal boundaries and have twisted periodic boundary conditions in the spatial directions: $\psi(x+L\hat i) = \exp (i\pi/5) \psi(x)$. 

\begin{figure}
\centering
\includegraphics[width=0.4\textwidth]{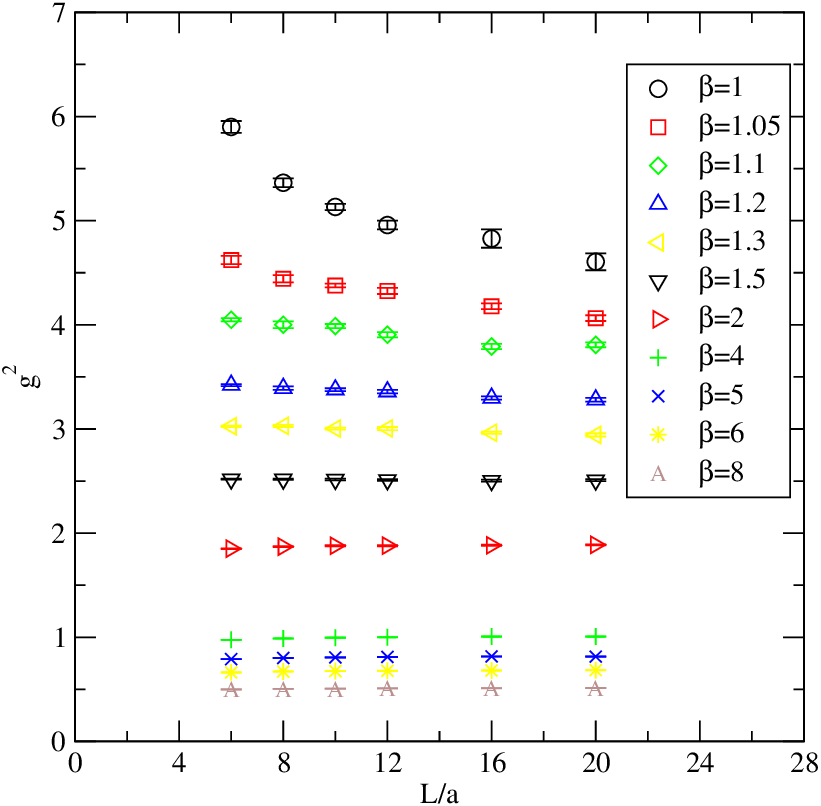}
\caption[a]{
  The measured values of the Schr\"odinger functional coupling
  $g^2(g_0^2,L/a)$ against $L/a$ at different $\beta_L$.
}
\label{fig:coupling}
\end{figure}

\begin{figure}
\centering
 \includegraphics[width=0.4\textwidth]{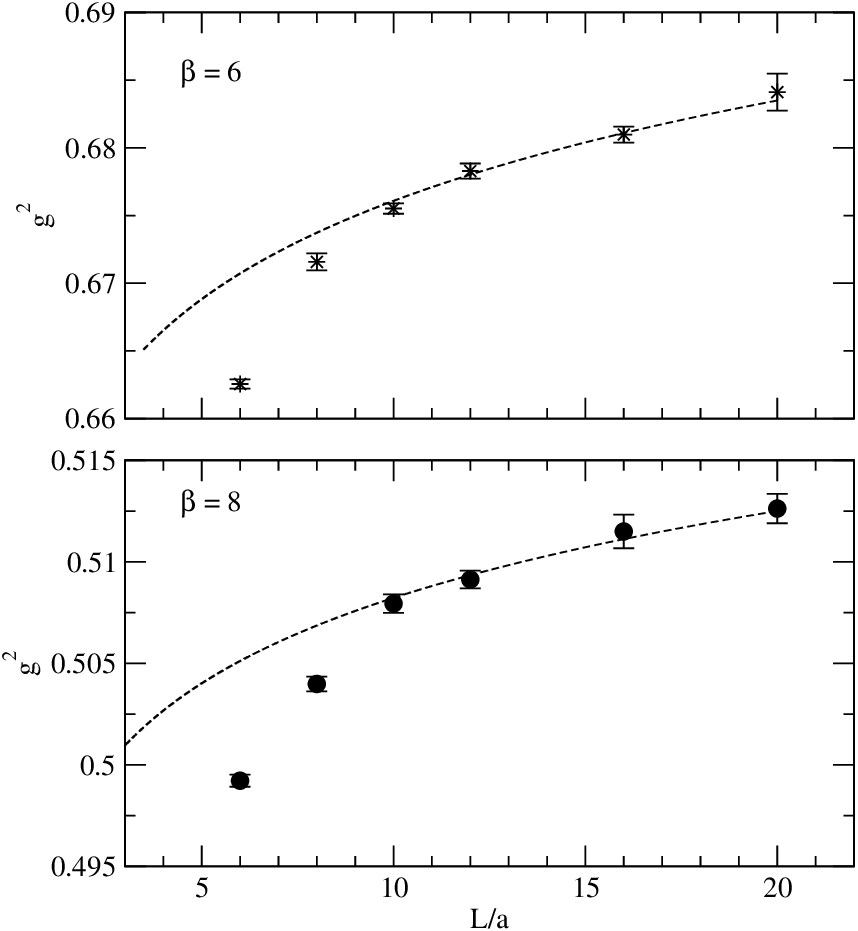}
\caption[a]{
  $g^2(g_0^2,L/a)$ against $L/a$ for $\beta_L=6$ and 8, compared against
  the running of the coupling in 2-loop perturbation theory (dashed lines).
}
\label{fig:zoom}
\end{figure}

The coupling constant is defined as the response of the system 
to the change of the background field:
\begin{align}
 \ev{\frac{\partial S}{\partial \eta}} = \frac{k}{g^2}.
\label{sfcoupling}
\end{align}
Here $k$ is a known function of $L/a$ and $\eta$ \cite{Luscher:1992ny}.
The measured values of $g^2(L/a,\beta_L)$ are given
in table \ref{table:coupling} and shown in Fig.~\ref{fig:coupling}.
In Fig.~\ref{fig:zoom} we zoom to the two smallest couplings (large $\beta_L$); it is clear that at large enough volumes ($L/a \gsim 10$) the points here reproduce perturbation theory \jr{ while smaller volumes deviate from it. }

\begin{table}
\include{couplings}
\caption{The measured values of $g^2$ at each $\beta_L=4/g_0^2$ and $L/a$.}
\label{table:coupling}
\end{table}

It has been shown that the boundary conditions (\ref{sfcoupling}) 
for adjoint SU(2) fermions generate rather large finite volume effects.
These can be reduced by 
halving the boundary angle to $\eta = 0.125\pi$
\cite{Karavirta:2012qd,Sint:2012ae,Hietanen:2014lha}.  
However, this reduces the magnitude of the background field and makes the measurement considerably noisier, and
thus we retain the boundary conditions in Eqs.~(\ref{sfboundary1}--\ref{sfboundary2}).

\begin{figure}
\centering
\includegraphics[height=0.45\textwidth]{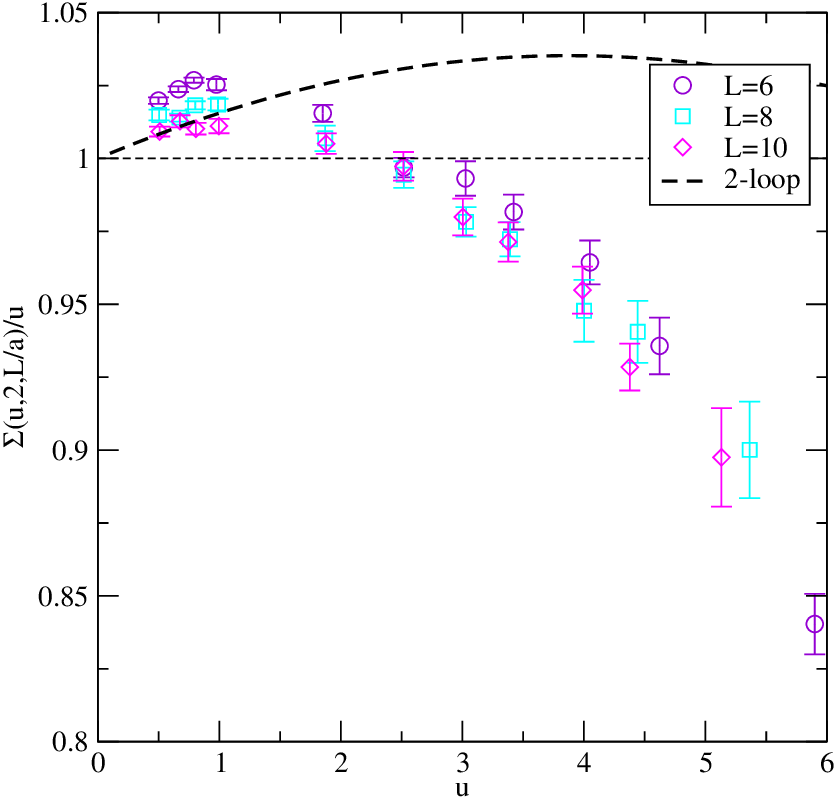}
\caption[b]{
  The scaled lattice step scaling function
  $ \Sigma(g^2,2,L/a)/g^2 = g^2(g_0^2,2L/a)/g^2(g_0^2,L/a) $ calculated 
  directly from the data in table \ref{table:coupling}.
  \kr{For comparison,} the black dashed line gives the continuum 2-loop perturbative result.
}
\label{fig:latstep}
\end{figure}

The running of the coupling is quantified by the step scaling function $\sigma(u,s)$, which
describes the change of the measured coupling when the linear size of the system is changed from $L$ to $sL$ while keeping the bare coupling $g_0^2$ constant \cite{Luscher:1993gh}:
\begin{align}
  \Sigma(u,s,L/a) &= g^2(g_0^2,sL/a)|_{u=g^2(g_0^2,L/a)} \\
  \sigma(u,s) &= \lim_{a/L \rightarrow 0} \Sigma(u,s,L/a)
\label{sigmaextrap}
\end{align}
We use $s=2$ and obtain the continuum limit from measurements at $L/a=6$, $8$ and $10$, pairing these with lattices with $L/a=12$, $16$ and $20$.  In Fig.~\ref{fig:latstep} we show the scaled step scaling function $\Sigma(u,2,L/a)/u$.  At weak coupling the largest volume measurements agree very well with the universal perturbative 2-loop result, but smaller volumes deviate from it significantly.  This can be understood from the behaviour of the measurements of the coupling in Fig.~\ref{fig:zoom}: from $L/a=10$ upwards the measurments are compatible with 2-loop perturbation theory, but \kr{at smaller volumes} there is significant deviation. \kr{At couplings $u \gsim 2$ this systematic difference between large and small volumes is not apparent.}
\kr{Nevertheless}, it is evident that the measurements already point towards a fixed point at around $g^2 \sim 2$--$3$.


The proper continuum extrapolation of the step scaling function in \eq{sigmaextrap} requires that the measurements at different $L/a$ and $2L/a$ -pairs are done at same value of $u=g^2(g_0^2,L/a)$.  However, for simplicity, the measurements of $g^2$ are done at a fixed set of bare couplings  $\beta_L = 4/g_0^2$.  We use here two different methods, {the widely used interpolation of the coupling $g^2(g_0^2,L/a)$ and a new method using a polynomial fit ansatz to step scaling, in order to enable taking the continuum limit.}

\subsection{{Interpolation of $g^2(g_0^2,L/2)$}}
The first 
method is based on interpolation of the measurements at each lattice size $L/a$ by fitting to a function of $g_0^2$.  We use here a rational interpolating function \cite{Karavirta:2011zg}
\begin{align}
 g^2(g_0^2,L/a) = g_0^2  \frac{ 1+\sum_{i=1}^n a_i g_0^2 }{ 1+\sum_{i=1}^m b_i g_0^2 } \,
 \label{eq:g2fitfunc}
\end{align}
with $n=m=3$.  These values were chosen to minimize the combined $\chi^2$ over degrees of freedom, calculated from the sum of $\chi^2$ and degrees of freedom for each lattice size.  The values of $\chi^2$ are given in table \ref{table:chi2}.
The stability of the interpolation is estimated by reducing $n$ or $m$ by $1$ and repeating the analysis.

%

\begin{table}
\centering
\begin{tabular}{ l l l l l l l l  }
\hline
\,
 $L/a$ & $6$ & $8$ & $10$ & $12$ & $16$ & $20$ & combined \\
\hline
 $\chi^2/d.o.f$ & $0.140$ & $0.863$ & $0.565$ & $0.381$ & $0.268$ & $0.738$ & $0.738$
 \\
\hline
\end{tabular}
\caption{The values of $\chi^2/d.o.f$ for each lattice size $L/a$. }
\label{table:chi2}
\end{table}

\begin{figure}
\centering
\includegraphics[height=0.45\textwidth]{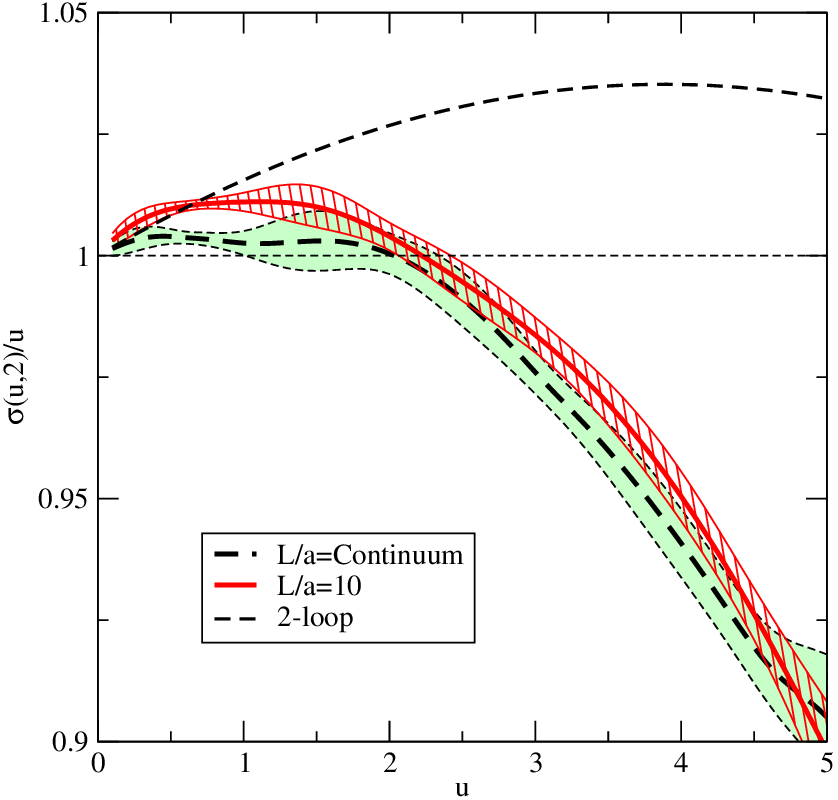}
\caption[b]{
  The scaled step scaling function $ \sigma(u,2)/u$, $u=g^2$, using only the largest
  volume pairs ($L/a = 10$ and $20$) (red hashed band) and with continuum extrapolation (green shaded band).
  The black dashed curve shows the universal 2-loop perturbative result.
}
\label{fig:Sigma}
\end{figure}

The interpolating function enables us to calculate the step scaling at any value of $u=g^2(g_0^2,L/a)$ within the interpolation range and enables us to obtain the continuum limit using the three $L/a$-values available.
We perform the continuum extrapolation by fitting the data 
to a function of the form
\begin{align}
\Sigma(u,2,L/a) =  \sigma(u,2) + c(u) \left ( L/a \right )^{-2}.
\label{continuum}
\end{align}
To propagate the error consistently throughout the analysis we divide the data into 40 jackknife blocks and perform the analysis separately on the blocks.
The final continuum extrapolated $\sigma(u,2)/u$ is shown in Fig.~\ref{fig:Sigma}, together with the 
step scaling function $\Sigma(u,2,10)$ obtained from the largest volume alone.  Due to the too large values of $\Sigma$ at small volumes and weak coupling, the continuum limit at small couplings deviates significantly from the perturbative value.  This deviation vanishes at $L/a \approx 10$, as is
evidenced by Fig.~\ref{fig:zoom}.  Therefore, we expect the $L/a=10$ result to be actually closer to the true continuum result than \kr{the} result from the extrapolation.  \kr{At couplings $u \gsim 2.5$ the continuum limit and $L/a=10$ result agree remarkably well.}

\kr{Because the lattices with $L/a < 10$ show significant finite volume effects at small coupling, it would be preferable to use only lattices larger than this in the continuum limit extrapolation.  In order to test this we have also analyzed the step scaling with a factor of $1.6$, using $L/a$-pairs $(10,16)$ and $(12.5,20)$.  The ``measurements'' at $L/a=12.5$ were synthesized from existing measurements, using either linear interpolation with $L/a=12$ and $16$ or quadratic interpolation using $L/a=10$, $12$ and $16$, with negligible differences.  While this method works in principle, in practice the lever-arm from $10$ to $12.5$ is so short that the continuum limit becomes very unstable and does not give a useful result.  In conclusion, a stable continuum limit would require simulations done at significantly larger volumes.}


The results indicate a fixed point close to $g^{*2}=2\ldots 3$. 
Using only $L/a=10$ results, the fixed point is at $g^{*2}=2.2(2)^{+0.6}_{-0.4}$, where the first error estimate gives the statistical error and the second includes estimated systematic error 
from the rational interpolation.  However, the continuum limit result tells us only that the fixed point is somewhere below
\kr{$g^2 \sim 3$, see Fig.~\ref{fig:Sigma}.}


\subsection{Power series ansatz}
\kr{The true $\beta$-function is a smooth function of $g^2$ and, at small coupling, its behaviour is determined by the perturbative part, which can be expanded in a power series of the coupling $g^2$.
This motivates us to 
try a different type of continuum extrapolation:
Because both $\sigma(u)$ and its discretization errors are smooth functions of $u$, we express them as truncated power series. This enables us to do a single fit to the step scaling data gathered at different couplings and lattice sizes.
Concretely, the fit function has the form
\begin{align} 
  &\sigma(u,2) =  1 + \sum_{i=1}^n c_i u^i \nonumber \\
  &\Sigma(u,2,a/L) = \sigma(u,2) + \sum_{k=2}^{n_a} f_k(u) \frac{a^k}{L^k} \label{eq.sigma_func_extrap}  \\
  &f_k(u) =  \sum_{l=1}^{m_k} c_{k,l} u^l, \nonumber
\end{align}
where $c_i$ and $c_{k,l}$ are fit parameters.  Because the discretization effects vanish as $u\rightarrow 0$, the expansion of $f_k$ starts at $u^1$.}

\kr{Due to the universality of the 2-loop $\beta$-function we know exactly the 
$u^0$, $u^1$ and $u^2$ -terms in the step scaling function $\sigma(u)$.
If the coefficients $c_1$ and $c_2$ are constrained to these universal values, we do not obtain an acceptable fit using only $O(a^2)$ discretization errors.
This should  not be surprising, considering the behaviour of the data at small couplings, as described in the previous section.}

\begin{figure}
\centering
\includegraphics[width=0.4\textwidth]{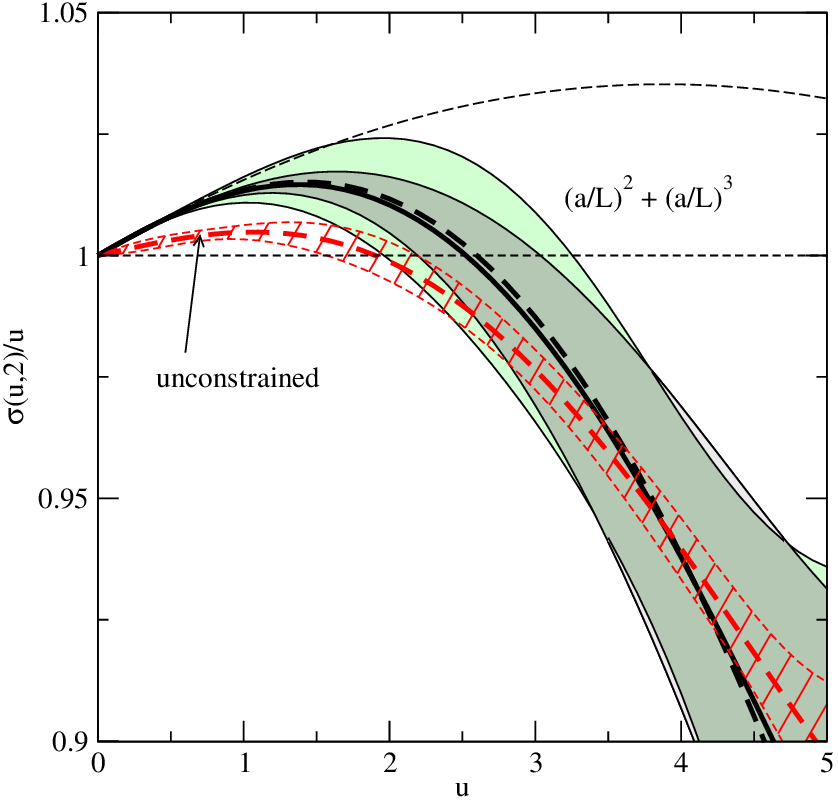}
\caption[b]{
  The continuum step scaling function resulting from extrapolations of the type in equation \ref{eq.sigma_func_extrap}. The shaded bands show the result when the fit is constrained by the universal $\beta$-function coefficients, 
\kr{where the wider band includes terms up to $u^5$, the narrower band up to 
$u^4$.}
The red hashed band shows the unconstrained result.  For details, see text.
}
\label{fig:sigma_fun_fit}
\end{figure}

\kr{However, the fitting procedure here allows us include also subleading $O(a^3)$ discretization effects.  When these are included we
%
%
obtain good and robust fits with varying number of fit parameters.  In Fig.~\ref{fig:sigma_fun_fit} we show two fits with 
$n_a=3$ (i.e. include $O(a^2)$ and $O(a^3)$ discretization errors): 
The first fit is done with $n=4$, $m_2=4$ and $m_3=2$, in total 8 parameters, with  $\chi^2/\mbox{d.o.f} \approx 20/25$.  The result is shown with a shaded band in Fig.~\ref{fig:sigma_fun_fit}.  In the second fit
we use  $n=5$, $m_2 = 5$ and $m_3=3$, all in all 11 parameters.
The resulting fit has $\chi^2/\mbox{d.o.f} \approx 18/22$, and is shown
with a broader shaded band in the figure.  The 
statistical error bands are obtained using jackknife analysis.  
The first fit is among the most constrained ones (i.e. least number of fit parameters) producing an acceptable result.  The second fit has more fit parameters and naturally produces a result which has wider statistical errors.  However, the good match of the fits supports the overall stability 
of the fitting procedure.}


\kr{By construction the above fits match the perturbative 2-loop result
perfectly at small $u$.}
If we do not constrain $c_1$ and $c_2$ to the known values 
but leave them as fit parameters, we obtain a 
result which is similar to the continuum limit obtained using the interpolation method, Fig.~\ref{fig:Sigma}.  In this case good $\chi^2$ ($\sim 19$ with 25 d.o.f) is obtained using only $O(a^2)$ discretization errors.  The resulting unconstrained curve is also shown in Fig.~\ref{fig:sigma_fun_fit}.  The error band is considerably narrower due to less freedom (missing $O(a^3)$ contribution) in the continuum extrapolation.  If $O(a^3)$ errors are included here, the error band becomes so broad that the fit loses its predictive power.

Thus, the advantage of the truncated power series fit is that it easily allows us to constrain the continuum limit with the known $\beta$-function behaviour. It is well controlled, enabling us to take into account some of the subleading discretization effects.  It also avoids the interpolation step, \eq{eq:g2fitfunc}.  The disadvantage is that the step scaling function 
is modelled with a truncated power series in $u$, in this case up to $u^4$ \kr{or $u^5$}.
However, we should keep in mind that the interpolating function, \eq{eq:g2fitfunc}, also restricts the structure of the resulting step scaling function.\footnote{%
  The use of the interpolating function can be avoided if the simulation parameters at different volumes are carefully tuned so that the measured couplings $u=g(g_0,L/a)$ are equal at each $L/a$.  In this case \eq{sigmaextrap} can be directly applied.  This was the procedure followed in the original Schr\"odinger functional analysis by Luscher  et al. \cite{Luscher:1993gh}.}
All in all, the result in Fig.~\ref{fig:sigma_fun_fit} is obtained using 8 parameters, whereas in the interpolation method $6\times 6 = 36$ fit parameters were used. 

Because the truncated series method gives more realistic behaviour at small couplings, we take \kr{our final estimate from the first fit shown in} Fig.~\ref{fig:sigma_fun_fit}.  Here the fixed point coupling is now in the interval $2.2 \lsim g^{*2} \lsim 3$, with a best fit value at $g^{*2} \approx 2.5$. This range agrees with earlier results in refs.~\cite{Hietanen:2009az,Bursa:2009we}; however, in~\cite{DeGrand:2011qd} a somewhat larger value $g^{*2} \approx 5$ (within the same scheme) is obtained.  In ref.~\cite{Rantaharju:2013bva} the fixed point was determined using the gradient flow, i.e. a different scheme, leading to result $g^{*2} \approx 5.5$.

\section{Anomalous dimension}
\label{gamma}

For the measurement of the  anomalous dimension of the mass, the spatial gauge links are set
to unity at temporal boundaries:
\begin{equation}
  U_i(\bs x,t=0) = U_i(\bs x,t=L) = \bs 1
\end{equation}
The mass anomalous dimension $\gamma$ is measured 
from the running of the pseudoscalar density renormalization constant
\cite{Capitani:1998mq,DellaMorte:2005kg}
\begin{align}
Z_P(L) = \frac{\sqrt{3 f_1} }{f_P(L/2)},
\label{Zp}
\end{align}
where the correlation function $f_P(t)$ is given in Eq.~(\ref{fp}) and is normalized 
using the boundary-to-boundary correlator
\begin{equation}
f_1 = \frac{-a^{12}}{3 L^{12}} \sum_{\bs u,\bs v, \bs y, \bs z} 
\ev{\bar \zeta'(\bs u)\gamma_5\frac12 \sigma^a \zeta'(\bs v)\,
  \bar\zeta(\bs y)\gamma_5\frac12 \sigma^a\zeta(\bs z)},
\end{equation}

Since the mass step scaling measurement is less noisy than the coupling measurement,
it is possible to use lattices of size $L=24$.
The measured values of $Z_P$ are given
in table \ref{table:zp}.

Now we can define the mass step scaling function as \cite{Capitani:1998mq}
\begin{align}
 \Sigma_P(u,s,L/a) &= 
    \left. \frac {Z_P(g_0,sL/a)}{Z_P(g_0,L/a)} \right |_{g^2(g_0,L/a)=u}
    \label{Sigmap}\\
 \sigma_P(u,s) &= \lim_{a/L\rightarrow 0} \Sigma_P(u,s,L/a).
\end{align}
We choose $s=2$ and find the continuum step scaling function $\sigma_P$ by
measuring $\Sigma_P$ at $L/a=6,8,10$ and $12$ and performing a quadratic extrapolation.

The mass anomalous dimension can then be obtained from the mass step scaling function \cite{DellaMorte:2005kg}.
Denoting the function estimating the anomalous dimension $\gamma(u)$ by
$\bar\gamma(u)$, we have
\begin{align}
   \bar\gamma(u) = -\frac{\log \sigma_P(u,s)}{\log s }.
 \label{eq:gammastar}
\end{align}
The estimator $\bar\gamma(g^2)$ is exact only at a fixed
point where $\beta(g^2)$ vanishes and deviates from the actual anomalous dimension when $\beta(g^2)$ is large.



\begin{table}
\include{zp}
\caption{The measured values of $Z_P$ at each $\beta_L$ and $L/a$.}
\label{table:zp}
\end{table}

\begin{figure*}
\centering
\includegraphics[height=0.4\textwidth]{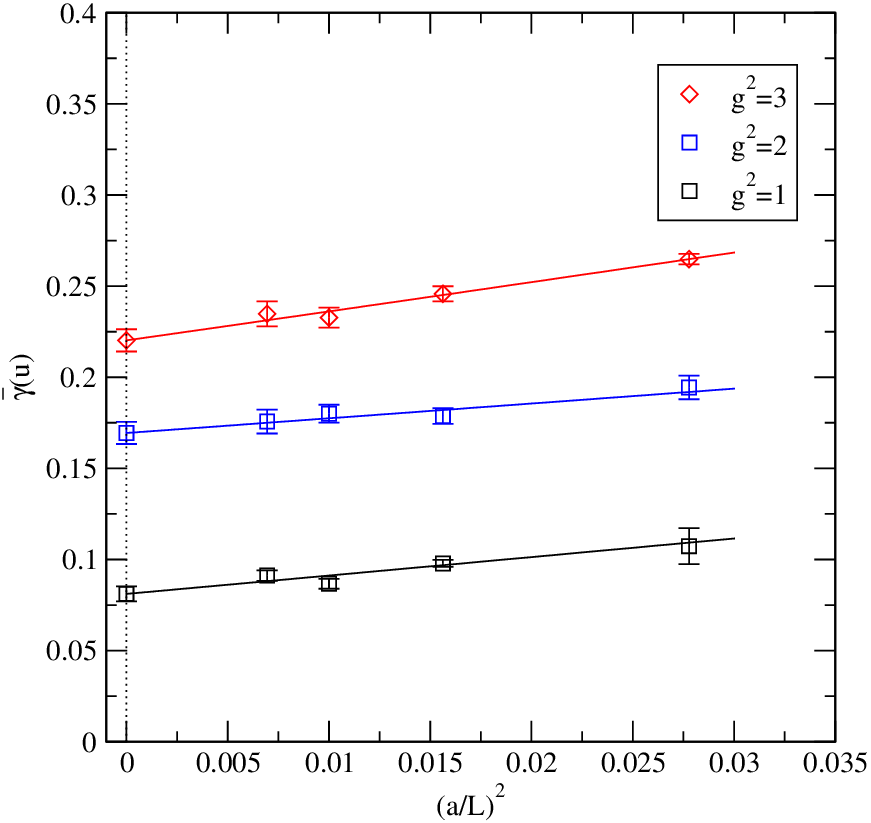}
\includegraphics[height=0.4\textwidth]{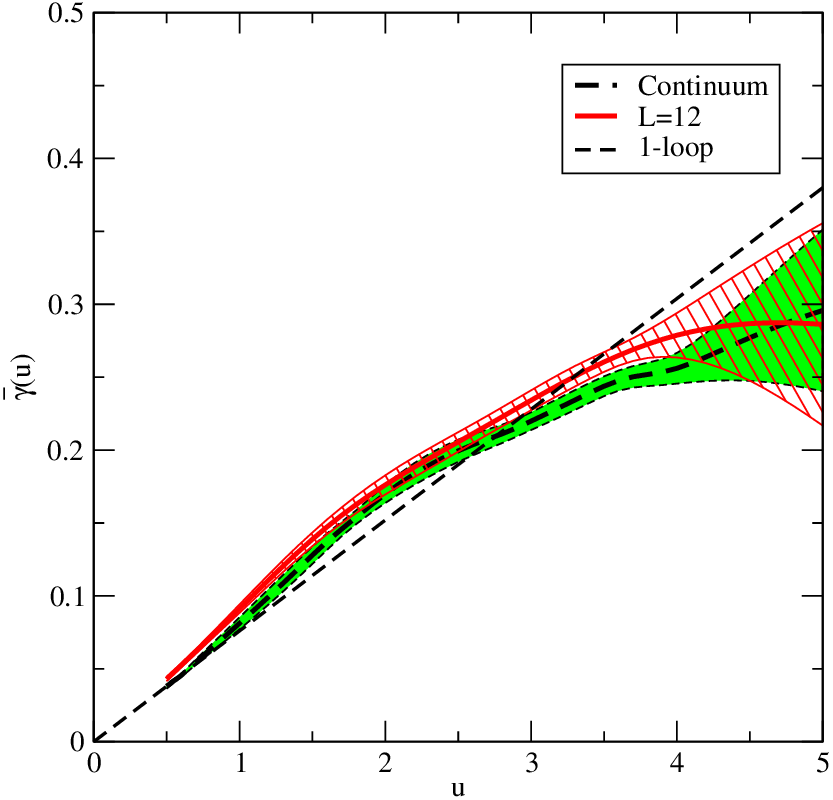}
\caption[b]{
  The estimate for the mass anomalous dimension $\bar\gamma(u)$.
  Left: Examples of the continuum extrapolation.
  Right: The green shaded band shows the continuum limit of $\bar\gamma$, and
  the red hashed band shows the result using only the largest lattice size $L/a=12$.
  The black dashed line gives the 1-loop perturbative result.
}
\label{fig:Gamma}
\end{figure*}

\begin{table}
\centering
\begin{tabular}{ l l l l l l l l l }
\hline
\,
 $L/a$ & $6$ & $8$ & $10$ & $12$ & $16$ & $20$ & $24$ & combined \\
\hline
 $\chi^2/d.o.f$ & $2.38$ & $0.526$ & $2.72$ & $0.425$ & $0.101$ & $0.238$ & $0.128$ & $0.930$
 \\
\hline
\end{tabular}
\caption{The values of $\chi^2/d.o.f$ for the interpolation of $Z_P$ for each lattice size $L/a$.}
\label{table:zpchi2}
\end{table}


For the analysis of the mass step scaling, we fit the data to an interpolating function. In this case a simple polynomial function is sufficient,
\begin{align}
  Z_P(\beta_L,L/a) = 1 + \sum_{i=1}^n
  c_i  g_0^{2i}. \label{eq:zpfunction}
\end{align}
where the optimal $\chi^2$ over degrees of freedom is given by $n=5$. The $\chi^2$ values for the fits are given in table \ref{table:zpchi2}.
The systematic error from this step is estimated by reducing $n$ by one and repeating the analysis.

We then calculate the mass step scaling function $\Sigma_P(u,s,L/a)$
in equation \eq{Sigmap} at $L/a=6,8,10$ and $12$.
The value for the coupling $u=g^2$ is obtained from the rational fit in \eq{eq:g2fitfunc}.
Finally, we calculate the estimating function $\bar\gamma(u,a/L)$ and find the continuum
limit $\bar\gamma(u)$ by fitting to a function of the form
$\bar\gamma(u,a/L) =\bar\gamma(u) + c(u) (a/L)^2$.
The result is shown in Fig.~\ref{fig:Sigma}.

At the fixed point we obtain the anomalous dimension \kr{$\gamma^* = 0.2 \pm 0.03$,}
where the dominant uncertainty comes from the location of the fixed point, $g^{*2} \approx 2.5^{+0.5}_{-0.3}$.  As can be seen from Fig.~\ref{fig:Sigma}, $\bar\gamma(u)$ is compatible with the perturbation theory within the range of $u$ studied.  However, in perturbation theory the IR fixed point typically happens at much larger coupling, and thus the anomalous dimension at the IRFP is correspondingly larger.

In ref.~\cite{DeGrand:2011qd} a larger result, $\gamma^* \approx 0.31(6)$, was obtained in the same scheme as used here.   In ref.~\cite{Patella:2012da} Patella used a different method to obtain again $\gamma^* \approx 0.37(2)$.  
\kr{In both of these cases the difference is in practice due to the larger value for the fixed point coupling.}



\section{Conclusions}
\label{checkout}

In this paper we have presented the results of a lattice study of the SU(2) gauge theory with two 
fermions in the adjoint representation of the gauge group. On the lattice the theory is 
formulated using a HEX smeared fermion action with tree level improvement and 
a partially smeared plaquette gauge action.
We expect this formulation to remove most of the $O(a)$ errors and to alleviate the higher order errors 
and allow us to \kr{investigate the continuum limit.}

We have measured the running coupling and the mass anomalous dimension 
in the Schr\"odinger functional scheme, using larger lattices than previous studies.  Our results confirm the existence of a non-trivial infrared fixed point.  The Schr\"odinger functional coupling at the fixed point is
$g^{\ast 2}\simeq 2.5^{+0.5}_{-0.3}$.  This agrees with the results in refs.~\cite{Hietanen:2009az,Bursa:2009we}, however, in these studies no proper continuum limit was possible.  De Grand et al.~\cite{DeGrand:2011qd} obtained $g^{\ast 2} \simeq 5$, a substantially larger value than us, although with a large uncertainty. In each of these studies different lattice actions were used. Therefore, while in the continuum limit all should give the same answer, at finite lattice spacings \kr{and without reliable continuum limit} the results may differ.  Indeed, as we  have observed here in Fig.~\ref{fig:zoom}, at $L/a \lsim 10$ the finite volume (equivalent to finite lattice spacing in Schr\"odinger functional scheme) effects remain substantial.  \kr{This makes the standard continuum limit extrapolation of the step scaling function unreliable.  We have also presented results from a continuum limit extrapolation using a truncated power series ansatz, which enables us to constrain the result with the universal 2-loop perturbative $\beta$-function coefficients.
Nevertheless, it may very well be that signficantly larger volumes are needed for a reliable continuum result. }

For the mass anomalous dimension at the fixed point we obtain \kr{$\gamma^\ast\simeq 0.2\pm 0.03$.}  Here the error is dominated by the uncertainty of the fixed point coupling $g^{\ast 2}$.  In general, $\gamma(u)$ follows the perturbative result closely up to $g^2 \approx 4$.

\acknowledgments 
This work is supported by the Academy of Finland
grants 267842, 134018 and 267286 and by the Danish National Research Foundation grant number DNRF:90.
JR acknowledges support from V\"ais\"al\"a foundation and
TR from the Magnus Ehrnrooth foundation.
The simulations were performed at the Finnish IT
Center for Science (CSC) in Espoo, Finland;
on the Fermi supercomputer at Cineca in Bologna, Italy,
under PRACE project 2012071257;
and on the k-computer at Riken AICS in Kobe, Japan. 

\end{document}

%% file: couplings.tex
\begin{tabular}{ l l l l l l l }
\hline
$ \beta_L$
 & $L/a=6$ & $L/a=8$ & $L/a=10$ \\
\hline
 $ 8 $ & $0.4992(3)$
& $0.5040(4)$
& $0.5079(5)$
 \\
 $ 6 $ & $0.6626(3)$
& $0.6716(6)$
& $0.6755(4)$
 \\
 $ 5 $ & $0.7900(5)$
& $0.8015(5)$
& $0.8066(6)$
 \\
 $ 4 $ & $0.9756(8)$
& $0.989(1)$
& $0.996(1)$
 \\
 $ 2 $ & $1.851(3)$
& $1.870(5)$
& $1.879(5)$
 \\
 $ 1.5 $ & $2.518(3)$
& $2.518(6)$
& $2.515(9)$
 \\
 $ 1.3 $ & $3.026(7)$
& $3.03(1)$
& $3.004(10)$
 \\
 $ 1.2 $ & $3.421(10)$
& $3.39(2)$
& $3.38(1)$
 \\
 $ 1.1 $ & $4.05(2)$
& $4.00(3)$
& $3.99(2)$
 \\
 $ 1.05 $ & $4.62(4)$
& $4.44(4)$
& $4.38(2)$
 \\
 $ 1 $ & $5.90(6)$
& $5.36(4)$
& $5.13(3)$
 \\
\hline
$ \beta_L$
 & $L/a=12$ & $L/a=16$ & $L/a=20$ \\
\hline
 $ 8 $ & $0.5091(4)$
& $0.5115(8)$
& $0.5126(7)$
 \\
 $ 6 $ & $0.6783(6)$
& $0.6810(6)$
& $0.684(1)$
 \\
 $ 5 $ & $0.8112(6)$
& $0.816(1)$
& $0.815(1)$
 \\
 $ 4 $ & $1.000(2)$
& $1.007(2)$
& $1.007(2)$
 \\
 $ 2 $ & $1.880(5)$
& $1.883(6)$
& $1.888(5)$
 \\
 $ 1.5 $ & $2.510(8)$
& $2.504(10)$
& $2.508(9)$
 \\
 $ 1.3 $ & $3.00(1)$
& $2.96(1)$
& $2.94(2)$
 \\
 $ 1.2 $ & $3.36(2)$
& $3.30(2)$
& $3.28(2)$
 \\
 $ 1.1 $ & $3.90(2)$
& $3.79(2)$
& $3.81(2)$
 \\
 $ 1.05 $ & $4.33(3)$
& $4.18(3)$
& $4.06(3)$
 \\
 $ 1 $ & $4.96(4)$
& $4.83(9)$
& $4.61(8)$
 \\
\hline
\end{tabular}

%% file: zp.tex
\begin{tabular}{ l l l l l l l l }
\hline
$ \beta_L$
 & $L/a=6$ & $L/a=8$ & $L/a=10$ & $L/a=12$ \\
\hline
 $ 8 $ & $0.9816(1)$
& $0.9615(2)$
& $0.9496(2)$
& $0.9404(3)$
 \\
 $ 4 $ & $0.9214(2)$
& $0.8908(4)$
& $0.8710(5)$
& $0.8565(8)$
 \\
 $ 2 $ & $0.7926(4)$
& $0.7475(6)$
& $0.7177(6)$
& $0.6982(10)$
 \\
 $ 1.5 $ & $0.7095(5)$
& $0.658(1)$
& $0.6254(7)$
& $0.603(1)$
 \\
 $ 1.3 $ & $0.6572(6)$
& $0.6014(9)$
& $0.5668(8)$
& $0.548(2)$
 \\
 $ 1.2 $ & $0.6222(5)$
& $0.568(1)$
& $0.537(1)$
& $0.510(2)$
 \\
 $ 1.1 $ & $0.5782(7)$
& $0.5262(9)$
& $0.4923(10)$
& $0.467(2)$
 \\
\hline
$ \beta_L$
 & $L/a=16$ & $L/a=20$ & $L/a=24$ \\
\hline
 $ 8 $ & $0.9289(6)$
& $0.9185(8)$
& $0.9123(9)$
 \\
 $ 4 $ & $0.833(1)$
& $0.820(1)$
& $0.804(2)$
 \\
 $ 2 $ & $0.665(2)$
& $0.640(3)$
& $0.623(4)$
 \\
 $ 1.5 $ & $0.566(2)$
& $0.537(3)$
& $0.522(3)$
 \\
 $ 1.3 $ & $0.508(2)$
& $0.485(2)$
& $0.465(4)$
 \\
 $ 1.2 $ & $0.472(2)$
& $0.451(3)$
& $0.429(3)$
 \\
 $ 1.1 $ & $0.434(2)$
& $0.407(2)$
& $0.385(3)$
 \\
\hline
\end{tabular}